\documentclass{article}

\usepackage[preprint, nonatbib]{nips_2018}

\usepackage{float}

\usepackage[square,numbers]{natbib}
\usepackage[numbers]{natbib}

\usepackage[utf8]{inputenc} 
\usepackage[T1]{fontenc}    
\usepackage{hyperref}       
\usepackage{url}            
\usepackage{booktabs}       
\usepackage{amsfonts}       
\usepackage{nicefrac}       
\usepackage{microtype}      
\usepackage{graphicx}
\usepackage{tabularx}
\usepackage{hyperref}
\usepackage{amsmath,amssymb,amsfonts}
\usepackage{enumitem}

\PassOptionsToPackage{off}{epstopdf}
\DeclareGraphicsExtensions{.pdf,.png,.jpg}

\title{Omni-QALAS: Optimized Multiparametric Imaging for Simultaneous $T_1$, $T_2$ and Myelin Water Mapping}

\author{
  Shizhuo Li \\
  \texttt{shizhuo0908@zju.edu.cn} \\
\And
 Unay Dorken Gallastegi \\
\And
Shohei Fujita \\
\And
 Yuting Chen \\
 \And
 Pengcheng Xu \\
 \And
 Yangsean Choi \\
 \And
 Huihui Ye \\
 \And
 Borjan Gagoski \\
 \And
 Huafeng Liu \\
 \And
 Berkin Bilgic \\
 \And
 Yohan Jun \\
  \texttt{yjun@mgh.harvard.edu} \\
}

\begin{document}

\maketitle

\begin{abstract}
Purpose: To improve the accuracy of multiparametric estimation, including myelin water fraction (MWF) quantification, and reduce scan time in 3D-QALAS by optimizing sequence parameters, using a self-supervised multilayer perceptron network.Methods: We jointly optimize flip angles, $T_2$ preparation durations, and sequence gaps for $T_1$ recovery using a self-supervised MLP trained to minimize a Cramér–Rao bound-based loss function, with explicit constraints on total scan time. The optimization targets white matter, gray matter, and myelin water tissues, and its performance was validated through simulation, phantom, and in vivo experiments.Results: Building on our previously proposed MWF-QALAS method for simultaneous MWF, $T_1$, and $T_2$ mapping, the optimized sequence reduces the number of readouts from six to five and achieves a scan time nearly one minute shorter, while also yielding higher $T_1$ and $T_2$ accuracy and improved MWF maps. This sequence enables simultaneous multiparametric quantification, including MWF, at 1 mm isotropic resolution within 3 minutes and 30 seconds.Conclusion: This study demonstrated that optimizing sequence parameters using a self-supervised MLP network improved $T_1$,  $T_2$ and MWF estimation accuracy, while reducing scan time. 
\end{abstract}

\section{INTRODUCTION}
Quantitative magnetic resonance imaging (qMRI) enables direct measurement of tissue physical parameters, such as $T_1$, $T_2$, $T_2$* relaxation times, myelin water fraction, and diffusion coefficients. Compared to traditional contrast-weighted imaging, qMRI offers the potential for improved standardization across systems and sites \cite{Weiskopf2013}, and provides more sensitive and specific detection of disease-related alterations \cite{Keenan2019,Tofts2003,Jara2022,Granziera2021,Fujita2021}. However, qMRI is often limited by long scan times, as exemplified by $T_1$ mapping using inversion recovery (IR) and $T_2$ mapping using multi-echo spin echo (MESE) sequences\cite{Deoni2005,Warntjes2008}. To address this, accelerated methods including DESPOT1\cite{Deoni2005}, MP2RAGE\cite{Marques2010}, Magnetic Resonance Fingerprinting (MRF)\cite{Ma2013}, MR multitasking \cite{Christodoulou2018} and echo planar time-resolved imaging (EPTI)\cite{Wang2019} have been developed, enabling rapid and simultaneous quantification of multiple parameters.

A notable example is 3D-QALAS (3D-Quantification using an interleaved Look-Locker Acquisition Sequence with $T_2$ preparation), which enables high-resolution whole-brain $T_1$ and $T_2$ mapping within clinically feasible scan times \cite{Kvernby2014,Kvernby2017}. It combines $T_2$ preparation, inversion pulses, and predefined flip angles across five FLASH readouts. Quantitative maps of $T_1$, $T_2$, and $PD$ are obtained via dictionary matching, though accuracy and efficiency remain limited.

Numerous efforts have aimed to reduce scan time via acceleration techniques such as compressed sensing \cite{Lustig2007,Feng2014}, wave-controlled aliasing in parallel imaging \cite{Bilgic2015}, and model-based deep learning reconstruction \cite{Kvernby2017,Hammernik2018}, which permit higher undersampling rates. Conventional QALAS reconstruction, however, assumes instantaneous k-space center acquisition, neglecting $T_1$ relaxation effects during the FLASH readout \cite{Zhu2022}. To address this, Jun et al. \cite{Jun2024,Tamir2017} introduced a subspace-based method that more accurately models relaxation dynamics, enhancing the fidelity of $T_1$ and $T_2$ maps. Beyond reconstruction-based methods, further improvement of QALAS is possible through sequence parameter optimization. Parameters such as flip angles (FAs), echo time (TE) of the $T_2$ preparation, and sequence gaps (GAPs) that allow $T_1$ recovery are typically set heuristically. Recent studies have applied Cramér–Rao Bound (CRB)-based optimization to improve mapping precision in MRF \cite{Zhao2019,Zhao2020,Crafts2022,Lee2019}, sodium MRF \cite{Kratzer2021}, ultra-short TE MRF \cite{Zhou2023}, and alternating FISP/PSIF sequences \cite{Li2024}. For QALAS, Arefeen et al. \cite{Arefeen2023} conducted preliminary FA optimization using CRB, though the optimization of other tunable parameters remain unexplored. 

Beyond $T_1$ and $T_2$, myelin water fraction (MWF) has gained attention as a biomarker of myelin integrity, with significant implications for understanding neurodevelopment and monitoring demyelinating diseases such as multiple sclerosis\cite{MacKay1994,Laule2006}. MWF quantifies the fraction of water trapped between myelin bilayers, which exhibits distinct relaxation properties (e.g., short $T_2$ values of 10–40~ms) compared to intra-/extracellular or free water\cite{Labadie2014}. While advanced techniques like MRF-based MWF mapping have demonstrated sensitivity to myelin changes in pediatric development\cite{Chen2019, Lancione2024}, the QALAS sequence, as originally designed, lacks adequate sensitivity in quantifying tissues with short relaxation times such as myelin water. This limitation arises from the use of a conventional TE=100~ms $T_2$ preparation module and a fixed flip angle of 4$^\circ$, which are suboptimal for capturing the rapid signal evolution of myelin water.

To address this issue, MWF-QALAS\cite{Fujita2025} was previously proposed, which introduced an additional, short-TE $T_2$ preparation module into the QALAS sequence. While this approach enhances the accuracy of short-$T_2$ signal quantification, it prolongs the acquisition by introducing an additional readout and employs heuristically determined sequence parameters, including two fixed $T_2$ preparation echo times of 20~ms and 80 ms.

We propose a CRB-based optimization framework using a self-supervised multilayer perceptron (MLP) network to optimize key sequence parametersof the MWF-QALAS, including FAs, TEs of the two $T_2$ preparation modules, and GAPs. This optimized sequence, which we refer to as Omni-QALAS (Optimized Myelin \& Neurometric Imaging), aims to improve the MWF quantification capability with a reduced number of FLASH readouts, thereby shortening the scan time while preserving $T_1$ and $T_2$ accuracy.

Our main contributions are as follows:

\begin{itemize}
    \item We firstly propose a CRB-based MLP optimization network for optimizing imaging sequence parameters, and demonstrate its application to QALAS and MWF-QALAS sequences. The MLP is capable of learning complex nonlinear relationships and can be easily extended to incorporate multi-objective loss functions, such as constraints on scan time. Compared to MWF-QALAS, our method requires shorter scan time (fewer FLASH readouts) while enabling more accurate $T_1$ and $T_2$ mapping, as well as improved MWF estimation. In vivo experiments demonstrate improved performance in terms of standard deviation.
    \item Using this framework, we derive the Omni-QALAS sequence by jointly optimizing FAs, GAPs, and TEs of the $T_2$ preparation, with explicit scan time regularization.
    \item We validate our method through ISMRM/NIST phantom experiments and in vivo studies, showing that Omni-QALAS achieves the best performance in MWF estimation while yielding better $T_1$ and $T_2$ accuracy than MWF-QALAS from a shorter acquisition.
    \item The resulting sequence and associated subspace reconstruction enables high fidelity $T_1$, $T_2$ and MWF mapping at 1 mm isotropic resolution in 3:30 min with whole brain coverage.
    \item All sequences were implemented using the open-source Pulseq framework \cite{Layton2017}; corresponding sequence files are available at the same repository.
\end{itemize}

\section{THEORY}
\subsection{Omni-QALAS Signal Model}
Omni-QALAS extends MWF-QALAS \cite{Fujita2025} to better capture short-$T_2$ components such as myelin water. As illustrated in Figure~\hyperref[overall]{\ref*{overall}(A)}, an additional $T_2$ preparation module with a short TE is inserted prior to the standard $T_2$ preparation. This design reduces the signal attenuation of myelin water during the preparation phase \cite{Nguyen2016}, thereby improving SNR and enhancing the stability of quantitative mapping.

During the period of $T_2$ sensitization achieved through an adiabatic $T_2$ preparation pulse, $T_2$ relaxation takes place, which can be expressed as follows:
\begin{equation}
M_{post}(\mathbf{r}) = M_{pre}(\mathbf{r}) e^{-\frac{\mathrm{TE}_{\mathrm{T_2prep}}}{T_{2}(\mathbf{r})}},
\end{equation}
where $M_{post}(\mathbf{r})$ represents the magnetization at location $\mathbf{r}$ after the $T_2$ preparation time $\mathrm{TE}_{\mathrm{T_2prep}}$, $M_{pre}(\mathbf{r})$ is the initial magnetization, and $T_{2}(\mathbf{r})$ is the $T_2$ relaxation time at location $\mathbf{r}$. Here, the $\mathrm{TE}_{\mathrm{T_2prep}}$ values are obtained through optimization.

After each $T_2$ preparation pulse, signal acquisition is performed using a FLASH readout. During the FLASH echo train, RF pulses are applied with flip angles $\alpha_i$, where $i$ denotes the $i^{th}$ excitation within the FLASH echo train, are obtained through optimization and vary over the echo train, whereas their initial values are set heuristically to $4^\circ$. \(\alpha_i\) is applied at each echo spacing \(\Delta \tau\).



 Based on this, the acquired transverse magnetization signal is expressed as:
\begin{equation}
S_{t+\Delta \tau}(\mathbf{r}) = \left\{ M_0^(\mathbf{r}) - \left( M_0^(\mathbf{r}) - M_t(\mathbf{r}) \right) e^{-\frac{\Delta \tau}{T_1^(\mathbf{r})}} \right\} \cdot \sin(\alpha_i).
\end{equation}


This is followed by a period of $T_1$ recovery gaps, during which $T_1$ relaxation occurs:
\begin{equation}
M_{t+GAP_i}(\mathbf{r})=M_{0}(\mathbf{r})-(M_{0}(\mathbf{r})-M_{t}(\mathbf{r}))e^{-\frac{GAP_i}{T_{1}(\mathbf{r})}},
\end{equation}

where $GAP_i$ is the $i$th $T_1$ recovery gap, as illustrated in Figure~\hyperref[overall]{\ref*{overall}(A)}, which is obtained through optimization. In the original QALAS sequence, GAP is derived from the time interval between readouts (originally set to 0.9~s) by subtracting the echo train length (ETL, defined as $ETL = n_{\text{echo}} \cdot \Delta \tau$, with $n_{\text{echo}} = 128$) and the duration of any preparation module (e.g., $T_2$ preparation or inversion pulse), if present.

There are many degrees of freedom associated with these variables, including FAs, TEs, and GAPs, which we propose to jointly optimize as detailed in the following section.

\subsection{CRB-based Optimization with Multilayer Perceptron Network}

\begin{figure*}[h]
\centering
\includegraphics[width=1\linewidth]{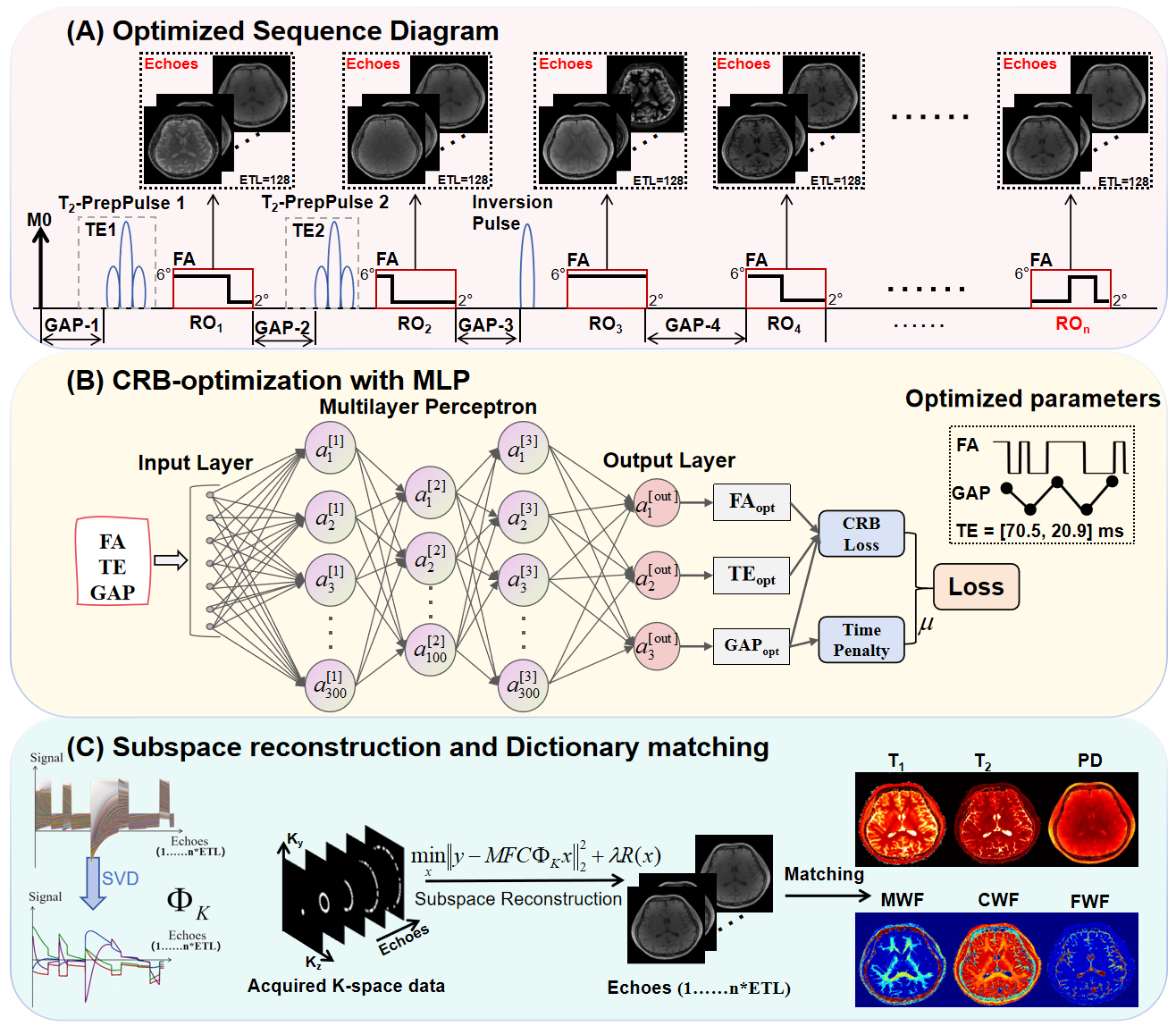}
\caption{\textbf{(A)} Omni-QALAS sequence diagram. An additional short-TE T\textsubscript{2} preparation pulse (T\textsubscript{2}-PrepPulse 0) is introduced before the QALAS readouts to enhance sensitivity to short-T\textsubscript{2} components such as myelin water. The flip angles (FAs) within each readout block are independently optimized and allowed to vary across time points. \textbf{(B)} Architecture of the self-supervised MLP network used for sequence optimization. The loss function consists of a CRB term and a time penalty term, balancing estimation accuracy and scan duration. The top-right inset shows the optimized FAs and GAPs. \textbf{(C)} Schematic of the subspace reconstruction. For each sequence, a separate dictionary was constructed, and the corresponding subspace bases were derived from the simulated signal evolutions using singular value decomposition (SVD). The quantitative parameter maps include T\textsubscript{1}, T\textsubscript{2}, PD, myelin water fraction (MWF), intra-/extra-cellular water fraction (CWF), and free water fraction (FWF).
}
\label{overall}
\end{figure*}
The Cramér-Rao Bound (CRB) is used to quantify the minimum variance of an unbiased estimator\cite{Aharon2024,Tang2023} and establishes a lower bound for the variance of such an estimator. Specifically, if \(\hat{\theta}\) is an unbiased estimator of the parameter \(\theta\), its variance satisfies:$\text{Var}(\hat{\theta}) \geq \frac{1}{I(\theta)}$, where $I$ represents the Fisher Information, which measures the amount of information that the observed data provides about the parameter \(\theta\):
\begin{equation}
\operatorname{Var}(\hat{\theta}) \geq V(\theta) = \left[ -\mathbb{E}\left[\frac{\partial^2 \ln \mathrm{p}(\mathrm{s} ; \theta)}{\partial \theta^2}\right] \right]^{-1} \text{,}
\end{equation}
where \(s\) is the observed QALAS sequence signal, ${\mathbb{E}}$ is the expectation, and $\theta$ represents the parameters to be optimized, such as \{$FA, TE, GAP$\}. $V(\theta)$ is the CRB matrix, which is equal to the inverse of Fischer Information Matrix (FIM). $\mathrm{p}$ represents the probability distribution of the observed data \(s\) given the parameter \(\theta\). $\ln \mathrm{p} $ is the log-likelihood. Under the Gaussian data model with identically distributed noise, the FIM can be simplified assuming Gaussian noise distributed as $\mathcal{N}(0, \sigma^2 \mathbf{I})$.

\begin{equation}
I(\theta) = \frac{1}{\sigma^2} \sum_n \left( \frac{\partial \mathbf{s}[n]}{\partial \theta} \right)^T \left( \frac{\partial \mathbf{s}[n]}{\partial \theta} \right)
\end{equation}
Therefore, we can optimize the parameters of the sequence by minimizing the CRB value, which can be expressed as:
\begin{equation}
\underset{\text{FA, TE, GAP}}{\operatorname{min}} V(\theta) \text{.}
\label{objective}
\end{equation}
When calculating the CRB values for different parameters, we use the weighting matrix $W=diag\left(\left[1/{T_2}^2,1/{T_1}^2,1/{PD}^2\right]\right)$ to normalize the CRB values\cite{Lee2019}. This helps prioritize tissues with shorter $T_1$ and $T_2$. We separately optimized the standard QALAS sequence for parenchymal $T_1$, $T_2$ mapping and the MWF-QALAS sequence for simultaneous MWF, $T_1$ and $T_2$ map estimation.  
We selected two representative tissue parameters: white matter = [$T_2$ = 70 \, \text{ms}, $T_1$ = 700 \, \text{ms}, $PD$ = 1] and gray matter = [$T_2$ = 80 \, \text{ms}, $T_1$ = 1300 \, \text{ms}, $PD$ = 1] for optimizing the QALAS sequence\cite{Zhao2020}. Additionally, we included myelin water = [$T_2$ = 20 \, \text{ms}, $T_1$ = 150 \, \text{ms}, $PD$ = 1] for optimizing the MWF-QALAS sequence. Furthermore, the optimization process tends to increase the $GAP$ value to achieve lower CRB values, which may lead to indefinitely long scan times. To constrain the scan time, we introduce an additional penalty term. The overall loss function is expressed as:
\begin{equation}
\text{Loss} = \sum_{i=1}^{\text{tissues}} W_i \cdot \text{CRB}_i + \mu \cdot \text{TimePenalty},
\label{crb}
\end{equation}
where, \( W_i \) denotes the weighting matrix for the \( i \)-th tissue type (white matter, gray matter, or myelin water). The first part calculates the CRB value of the signal using the optimized parameters, while the second part calculates the $TR$ as a time penalty term to constrain the sequence duration. The parameter $\mu$ was selected according to Figure~S1 to balance the CRB value and scan time.

In order to solve this non-linear and non-convex objective, we propose using an MLP network. We compared different MLP architectures through ablation experiments (see Figure~S2) and ultimately selected a structure of $300 \times 100 \times 300$, as shown in Figure~\hyperref[overall]{\ref*{overall}(B)}.

\subsection{Reconstruction}
The optimized FAs vary dynamically across the acquisition, which exacerbates artifacts introduced by conventional reconstruction methods due to the more complicated nature of tissue relaxation \cite{Zhu2022}. To address this issue, we adopt the subspace reconstruction approach proposed by Jun et al. \cite{Jun2024,Tamir2017}, which can be expressed as:
\begin{equation}
\min_{x} \| y - Ax \|_2^2 + \lambda R(x)\
\end{equation}

where, $A = MFC \Phi_k$, $R(x) = \| \Psi x \|_1$, $x$ represents the subspace coefficients to be reconstructed. $M$ are the k-space sampling mask, $F$ is the Fourier transform, $C$ is the multi-coil sensitivity map and $\Psi$ is the wavelet transform. The subspace basis $\Phi_k$ is obtained by performing SVD on the simulated dictionary. $K$ represents the selection of the first $K$ principal components. $\lambda$ is the regularization parameter that controls the balance between the two terms. 

The MWF quantification was performed using a dictionary-based partial volume decomposition approach\cite{Chen2019,Weigel2015}. We estimate four physiologically distinct compartments: (1) myelin water (MW), (2) white matter (WM), (3) gray matter (GM), and (4) free water (FW). Following the approach described in Labadie et al.\cite{Labadie2014}, these components were defined during the construction of the fraction dictionary as follows:

\begin{equation}
\begin{aligned}
&\text{MW:}~(T_1, T_2) = (150~\text{ms}, 20~\text{ms}) \\
&\text{WM:}~(T_1, T_2) = (1100~\text{ms}, 70~\text{ms}) \\
&\text{GM:}~(T_1, T_2) = (1300~\text{ms}, 80~\text{ms}) \\
&\text{FW:}~(T_1, T_2) = (4500~\text{ms}, 500~\text{ms}) \\
\end{aligned}
\end{equation}

The partial volume dictionary was constructed through linear combinations of normalized basis signals, where the partial volume fractions were discretized from 0 to 1. The final parametric maps were generated as:




\begin{equation}
\text{MWF} = P_{MW}, \quad \text{CWF} = P_{WM}+P_{GM}, \quad \text{FWF} = P_{FW}
\end{equation}
Where $P$ is partial volume fraction. CWF is cellular water fraction. Using WM and GM as two separate compartments, along with the Figure~S3, demonstrates improved performance compared to using a single combined intra-/extra-cellular water compartment, as it helps prevent gray matter signal from leaking into the free water component. 

\section{METHOD}

\subsection{Implementation details}  

In network optimization, the initial sequence parameters were set heuristically as follows: $\text{FA} = 4^\circ$, $\text{TE} = [20, 80] \, \text{ms}$, and $\text{GAPs}$ = [135.58, 75.58, 58.93, 165.28, 165.28]~ms, which served as the input to the network. Here, TE corresponds to the durations of two separate $T_2$ preparation pulses. The output of the network consists of the optimized $FA$, $TE$, and $GAP$. 

The optimized flip angles exhibit an alternating pattern between its maximum and minimum values. We constrain the parameter ranges as follows: The range \( 2^\circ \leq \mathrm{FA}\leq 6^\circ \) was chosen to balance the alternating pattern. TEs were constrained as \( 10\,\text{ms} \leq \mathrm{TE} \leq 200\,\text{ms} \). In addition to the time penalty consideration, we constrain $0\,\text{ms} \leq \mathrm{GAP} \leq 250\,\text{ms}$. To further reduce the acquisition time, the number of readouts was decreased from 6 (as in the existing MWF-QALAS) to 5. 

In terms of DIP reconstructions, We set $K = 4$ and $\lambda = 2 \times 10^{-6}$ in our experiments, following the subspace approach\cite{Jun2024}. 

To perform signal matching, dictionaries were constructed with discretized $T_1$,  $T_2$ and $B_1^+$ values in the ranges of [50, 5000] ms, [5, 2000] ms, [0.65, 1.35] and [0.6, 1.0] respectively. A dictionary mapping combinations of $T_1$ and $T_2$ to simulated inversion efficiencies(IE) was generated using Bloch equation simulations, enabling accurate parameter estimation in reconstructions.

In this study, we compare three sequences: the original QALAS \cite{Kvernby2014}., MWF-QALAS\cite {Fujita2025}, and the proposed Omni-QALAS, which is an optimized extension of MWF-QALAS.

\subsection{Simulation Experiments}
We constructed a digital phantom, as depicted in Figure~\ref{simulation}, with $T_1$ values ranging from [0, 3000 ms] and $T_2$ values ranging from [0, 300 ms]. Noise with SNR levels of 20, 12, and 8 was added to the phantom. We define the signal-to-noise ratio as: $SNR = (s/\sigma)^2$, where, $\sigma$ denotes the pre-specified noise level and $s$ denotes the average value of the k-space. 

Signal simulations and matching for the sequences were performed using the Bloch equations. To compare the quantitative accuracy of different sequences for various tissues, we selected three regions in Figure~\ref{simulation}, representing MW, GM, and WM, for evaluation.

\subsection{Phantom Experiments}
\label{phantom}
Phantom experiments were conducted on an ISMRM/NIST system phantom using a 3T MAGNETOM Prisma scanner equipped with a 32-channel (32ch) head receive array. The reference $T_1$ and $T_2$ maps were acquired using IR-FSE and SE-FSE sequences, respectively. For $T_1$ mapping, IR-FSE utilized inversion times of [25, 100, 200, 500, 1000, 2000, 3500] ms, and used Relaxation-Driven Non-Linear Least Squares method with phase removal \cite{barral2009robust}. For $T_2$ mapping, SE-FSE employed echo times of [10, 30, 50, 70, 90, 150, 250, 400] ms, and used the Variable Projection method \cite{zhao2018variable,golub1973differentiation}. During the acquisition, an acceleration factor of R=5 using variable-density Poisson-disc sampling was applied. 

In the phantom, the 14 spheres were divided into two groups: the outer 8 spheres exhibit relatively large $T_1$ and $T_2$ values, resembling those of gray and white matter, while the inner 6 spheres have smaller $T_1$ and $T_2$ values, similar to those of myelin water. 

\subsection{In vivo Experiments}

This experiment was conducted on healthy volunteers with written consent on a 3T MAGNETOM Prisma scanner (Siemens Healthineers, Erlangen, Germany) with a 32-channel (32ch) head receive array. The scan parameters for all sequences are summarized as follows. Omni-QALAS used two T\textsubscript{2}-prep echo times (70.9~ms and 20.8~ms), five readouts with GAPs of [156.48,4.08,149.93,15.28,265.28]~ms, and variable flip angles (shown in Figure~\ref{overall}). MWF-QALAS used two  T\textsubscript{2}-prep echo times (20~ms and 80~ms), fixed flip angles of $4^\circ$, six readouts with [135.58, 75.58, 58.93, 165.28, 165.28, 165.28]~ms GAPs. QALAS used one  T\textsubscript{2}-prep echo time (100~ms), fixed flip angles of $4^\circ$, five readouts with [55.58, 58.93, 165.28, 165.28, 165.28]~ms GAPs. B$_1^+$ maps were acquired using a vendor-provided turbo-FLASH sequence\cite{Chung2010} for subsequent B$_1^+$ inhomogeneity correction. The maps were interpolated to match the matrix size of the 3D-QALAS images, and their values were clipped to fall within the range defined by the simulation dictionary. 

The reference $T_1$ and $T_2$ maps were acquired using IR-FSE and SE-FSE scans, respectively, with the same setup as described in Section~\ref{phantom}.  For the reference MWF, a 3D-GRASE sequence was used to acquire 32 echo signals \cite{Piredda2021}. The acquisition parameters were as follows: voxel size = 1.0~$\times$~1.0~$\times$~3.0~mm$^3$, TR = 1250.0~ms, and 32 echo times (TEs) ranging from 12.22~ms to 391.04~ms with a uniform echo spacing of 12.22~ms. The reference MWF estimation was performed using the multi-exponential T$_2$ analysis method proposed by Prasloski et al.~\cite{Prasloski2012}, where the MWF was defined as the ratio of signal components with T$_2$ values between 10~ms and 40~ms to the total T$_2$ signal.

\subsubsection{Five-minute Omni-QALAS for 1~mm\textsuperscript{3} $T_1$, $T_2$ and MWF mapping}
In this experiment, the sequences used an isotropic voxel size of 1 mm\textsuperscript{3}, field of view (FOV) = 256~$\times$~240~$\times$~208~mm\textsuperscript{3}; matrix size = 256~$\times$~240~$\times$~208 and acceleration factor = 5 using variable-density Poisson-disc sampling.

The scan time for QALAS was 5 minutes, for MWF-QALAS was 5 minutes and 50 seconds, and for Omni-QALAS was 5 minutes.

\subsubsection{Higher Undersampling Rate and Spatial Resolution in Omni-QALAS}
In this experiment, we evaluated three Omni-QALAS sequences: \textbf{Omni-R7-1mm\textsuperscript{3}}, \textbf{Omni-R5-0.79mm\textsuperscript{3}}, and \textbf{Omni-R7-0.79mm\textsuperscript{3}}.

Omni-R7-1mm\textsuperscript{3} used a voxel size of 1 mm isotropic with a 7-fold undersampling rate using variable-density Poisson-disc sampling, achieving a scan time of 3 minutes and 30 seconds.

Omni-R5-0.79mm\textsuperscript{3} and Omni-R7-0.79mm\textsuperscript{3} both used a voxel size of 0.79 mm isotropic, with a field of view (FOV) of 256~$\times$~240~$\times$~208~mm\textsuperscript{3} and a matrix size of 324~$\times$~288~$\times$~256. These sequences used 5-fold and 7-fold undersampling rates, resulting in scan times of 7 minutes and 5 minutes, respectively.

\section{RESULTS}

\subsection{CRB comparision}
Figure~S4 presents the $T_1$, $T_2$ CRB loss values for WM, GM, MW, and the overall signal across the three sequences. The CRB loss is calculated based on the first term in Equation~\ref{crb}.  Notably, the QALAS sequence exhibits a substantially high $T_2$ CRB value for myelin water, reaching nearly 20{,}000, which leads to poor quantification accuracy for short $T_2$ components. This issue is significantly alleviated by introducing a $T_2$ preparation pulse targeting short $T_2$ signals, which effectively reduces the CRB loss in myelin water. Furthermore, Omni-QALAS substantially reduces CRB values for all tissues. Specifically, the overall CRB for $T_1$ is reduced by 45.8\%, with a 46.1\% reduction observed for myelin water. For $T_2$, the overall CRB decreases by 42.5\%, with a 53.2\% reduction for myelin water.

\subsection{Simulation}

\begin{figure*}[h]
\centering
\includegraphics[width=1\linewidth]{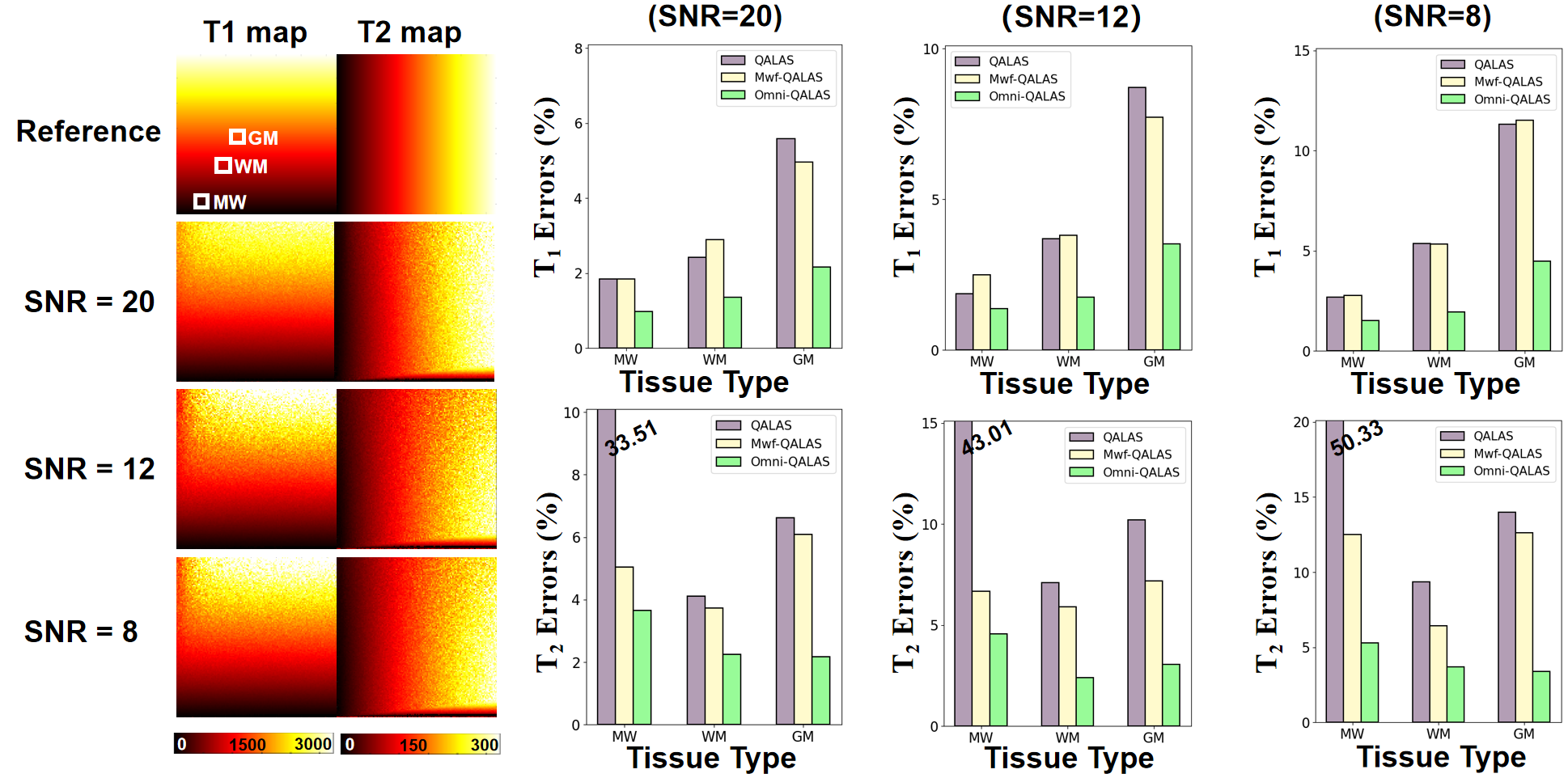}
\caption{The figure shows the simulation results. Three different levels of noise were added to the reference signals. The results of Omni-QALAS are displayed on the left side. Three regions of interest (ROIs), corresponding to myelin water (MW), white matter (WM), and gray matter (GM), were selected to quantitatively compare the estimation accuracy of the three sequences. The relative quantification errors under different noise levels are shown on the right.
}
\label{simulation}
\end{figure*}

The results of Omni-QALAS under different noise levels are shown on the left side of Figure~\ref{simulation}. The right side of the figure displays the errors relative to the reference for each sequence under different noise levels for each tissue type. We define the relative errors as: $error=\|\mathbf{I}-\hat{\mathbf{I}}\|_2 /\|\mathbf{I}\|_2$, where $\mathbf{I}$ and $\hat{\mathbf{I}}$ respectively denote the reference parameter map and reconstructed parameter map.

The results indicate that the QALAS sequence shows substantial $T_2$ quantification errors in the myelin water region, which corresponds to the highest CRB loss observed in the CRB comparison experiment for myelin water. However, introducing an additional $T_2$ preparation pulse (MWF-QALAS) significantly improves the quantification accuracy. Notably, while MWF-QALAS enhances the quantification accuracy for myelin water, it shows comparable performance to QALAS in gray and white matter regions. Notably, MWF-QALAS exhibits slightly worse performance in $T_1$ quantification. Omni-QALAS improves both $T_1$ and $T_2$ quantification. For $T_1$, it achieves 1.7-, 2.2-, and 2.5-fold improvements in MW, WM, and GM, respectively, compared to QALAS, and 1.8-, 2.4-, and 2.4-fold improvements compared to MWF-QALAS. For $T_2$, similar improvements of $\sim$9.4-, 2.5-, and 3.5-fold over QALAS and $\sim$1.7-, 2.0-, and 3.0-fold over MWF-QALAS are observed in MW, WM, and GM, respectively.

\subsection{NIST Phantom}

\begin{figure*}[htbp]
  \centering
  \includegraphics[width=1\linewidth]{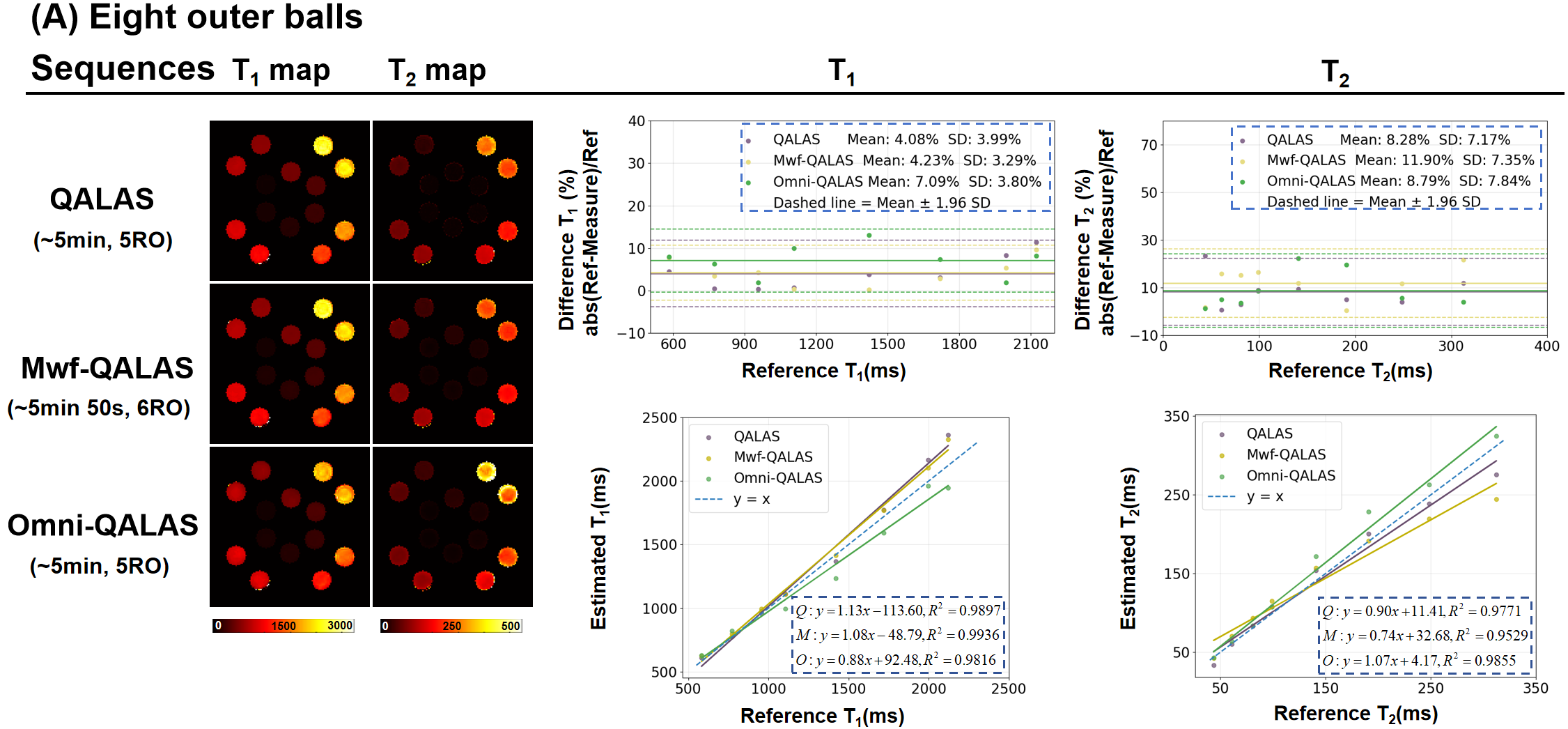}

  \includegraphics[width=1\linewidth]{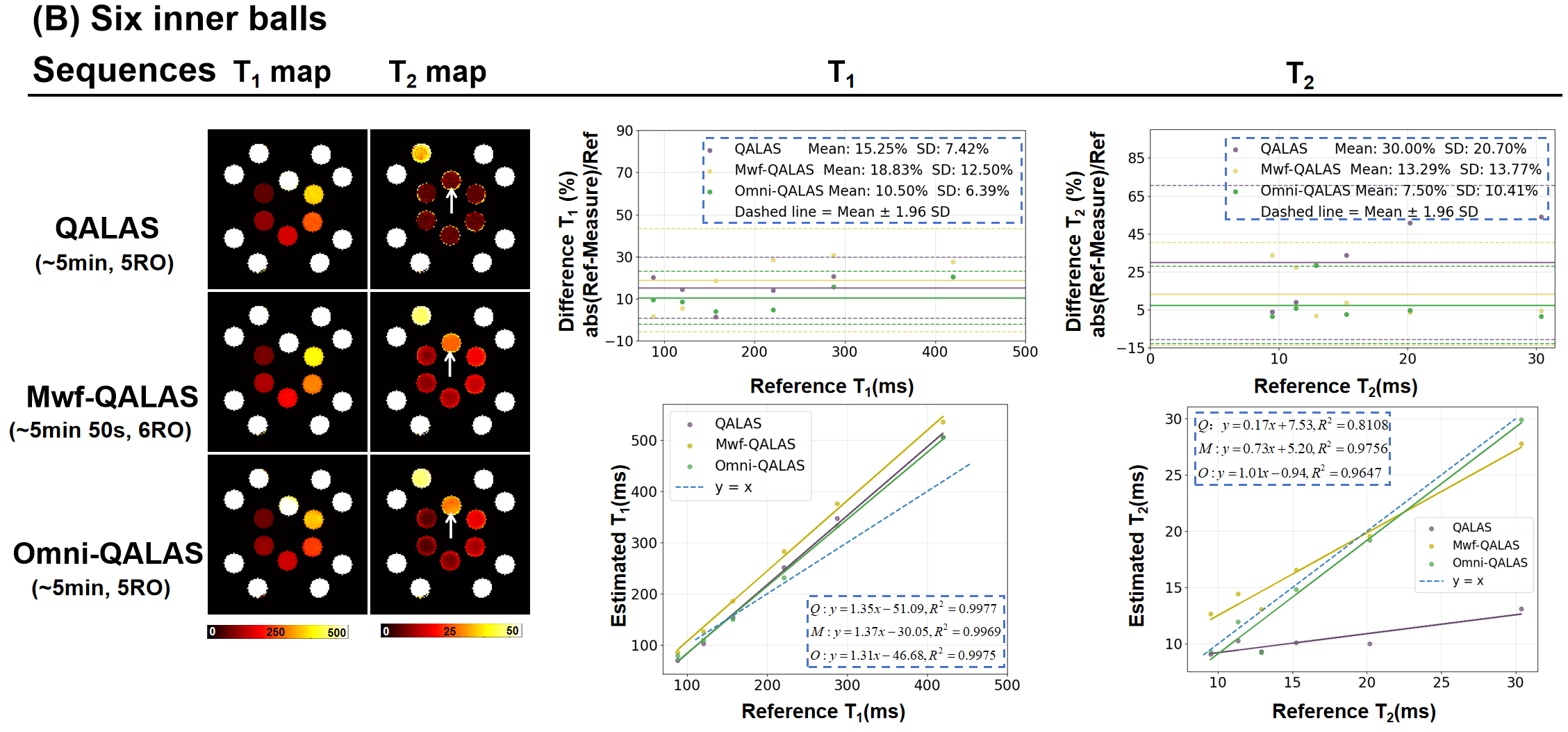}
  \caption{The figure presents the results of three sequences acquired on the NIST phantom. The top row focuses on the eight outer spheres, while the bottom row shows results for the six inner spheres. The right panel displays Bland–Altman plots and linear regression fits.}
  \label{nist}
\end{figure*}

The experimental results are shown in Figure~\ref{nist}. To better visualize these two groups, two separate colorbars were used in the figure. the upper one for the outer 8 spheres and the lower one for the inner 6 spheres. On the right, Bland–Altman plots and linear regression plots for each sequence are shown.

For the outer 8 spheres, the QALAS sequence demonstrated the best performance, with mean errors of 4.08\% for $T_1$ and 8.28\% for $T_2$, outperforming both MWF-QALAS and Omni-QALAS. Nevertheless, Omni-QALAS still showed improved $T_2$ quantification compared to MWF-QALAS, with the regression slope increasing from 0.74 (MWF-QALAS) to 1.07 (Omni-QALAS).

In contrast, for the inner 6 spheres, QALAS exhibited poor quantification accuracy, particularly for $T_2$, failing to reliably quantify or differentiate the spheres. This limitation was partially addressed by the additional $T_2$ preparation pulse in the MWF-QALAS sequence, which improved the regression slope to 0.79. Omni-QALAS further improved the slope to 1.01 while also reducing the scan time. Overall, Omni-QALAS significantly enhanced quantification accuracy for tissues with short $T_2$ values (e.g., myelin water), without requiring additional readouts. Moreover, it achieved comparable $T_1$ and $T_2$ quantification in gray and white matter relative to the original QALAS.

\subsection{In vivo}

\begin{figure*}[h]
\centering
\includegraphics[width=1\linewidth]{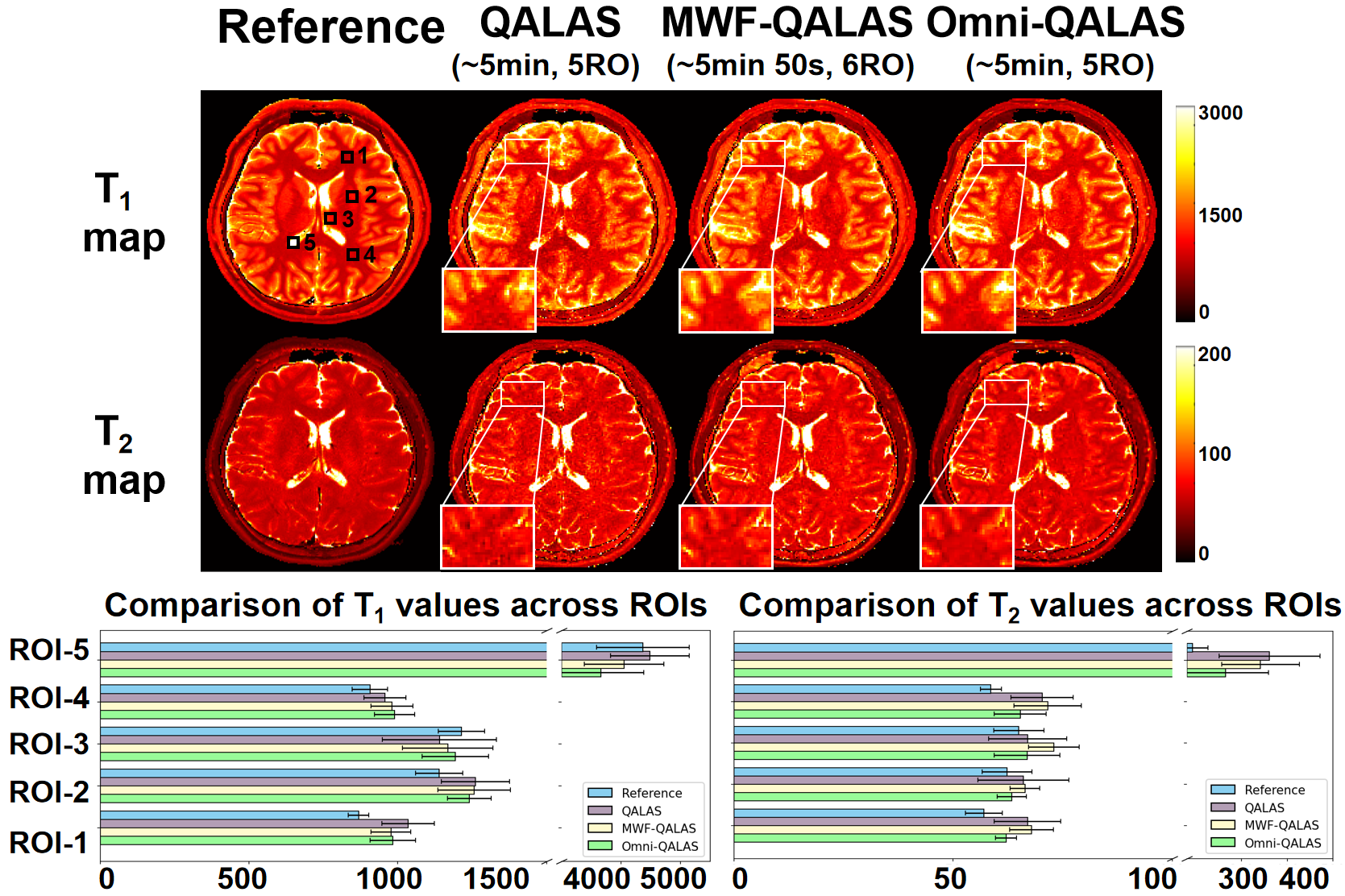}
\caption{The figure shows the results obtained from a healthy volunteer. The upper panel displays the T\textsubscript{1} and T\textsubscript{2} maps obtained using different sequences. Five regions of interest (ROIs) were selected for comparison across sequences. The mean and standard deviation of the quantitative values within these ROIs are shown in the lower panel.
}
\label{t1t2}
\end{figure*}

\begin{figure*}[h]
\centering
\includegraphics[width=1\linewidth]{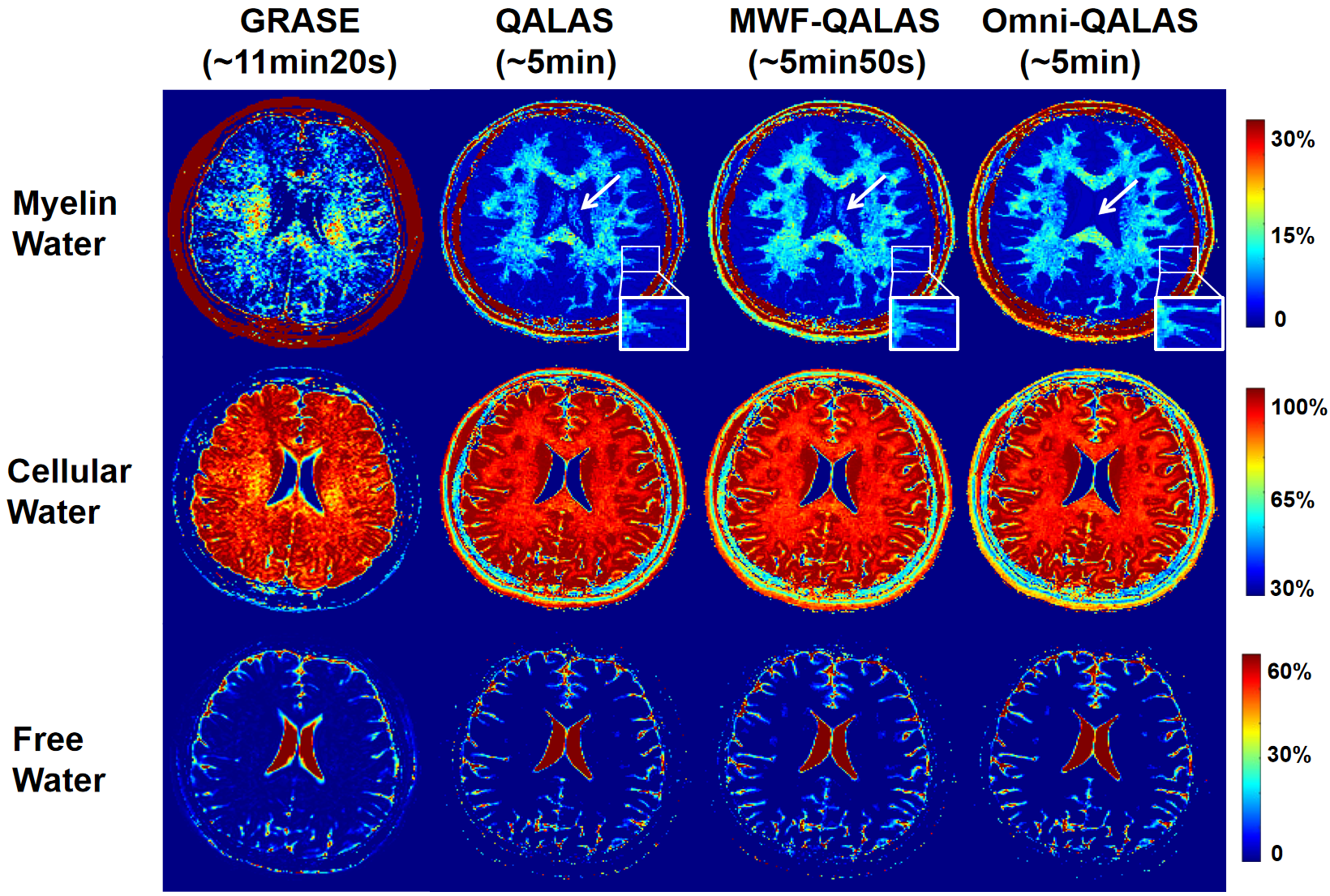}
\caption{The figure illustrates the experimental results of water fraction mapping from a healthy volunteer. Three types of water fractions are shown: myelin water, intra/extracellular water, and free water.
}
\label{mwf}
\end{figure*}

\begin{figure*}[htbp]
\centering
\includegraphics[width=1\linewidth]{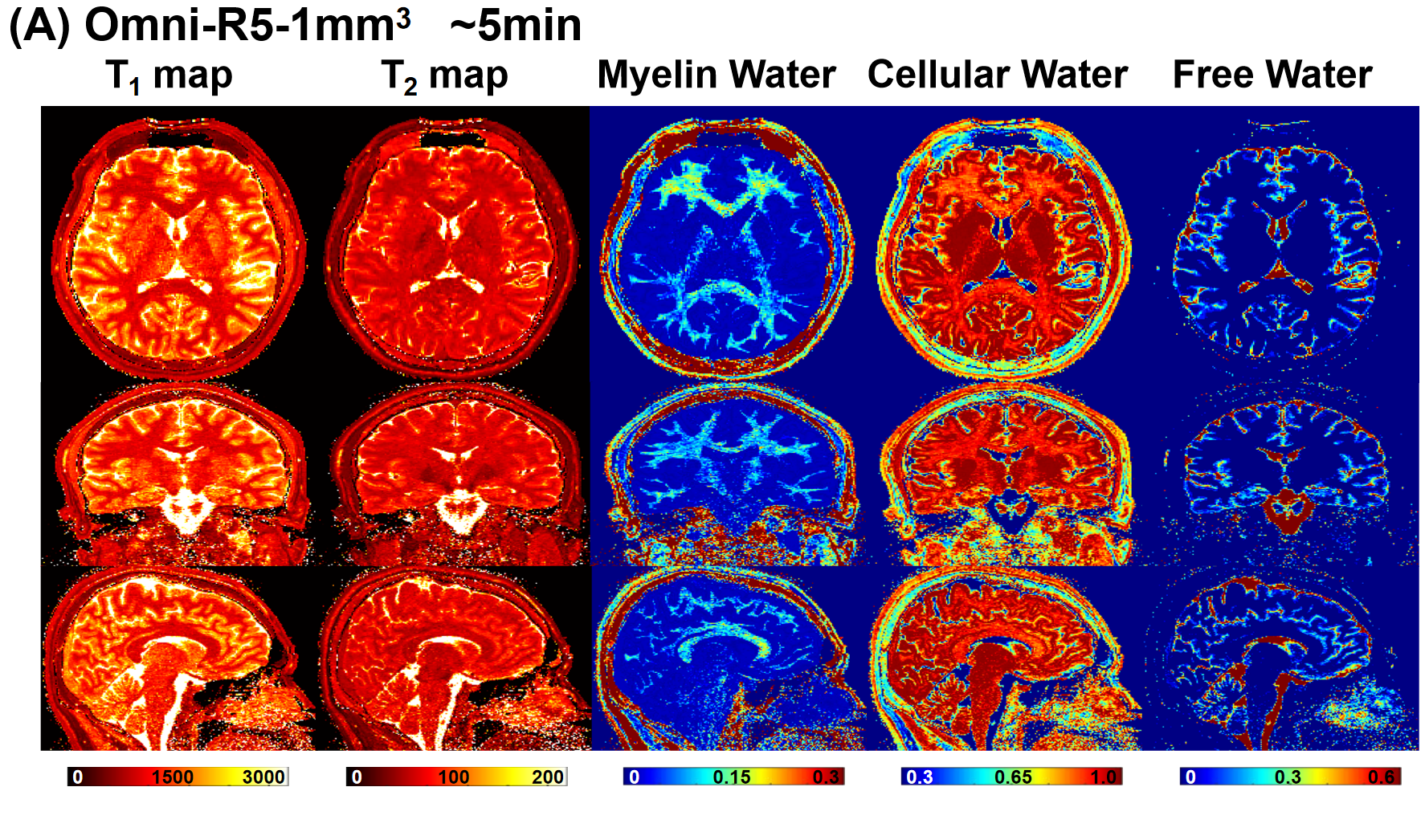}

\includegraphics[width=1\linewidth]{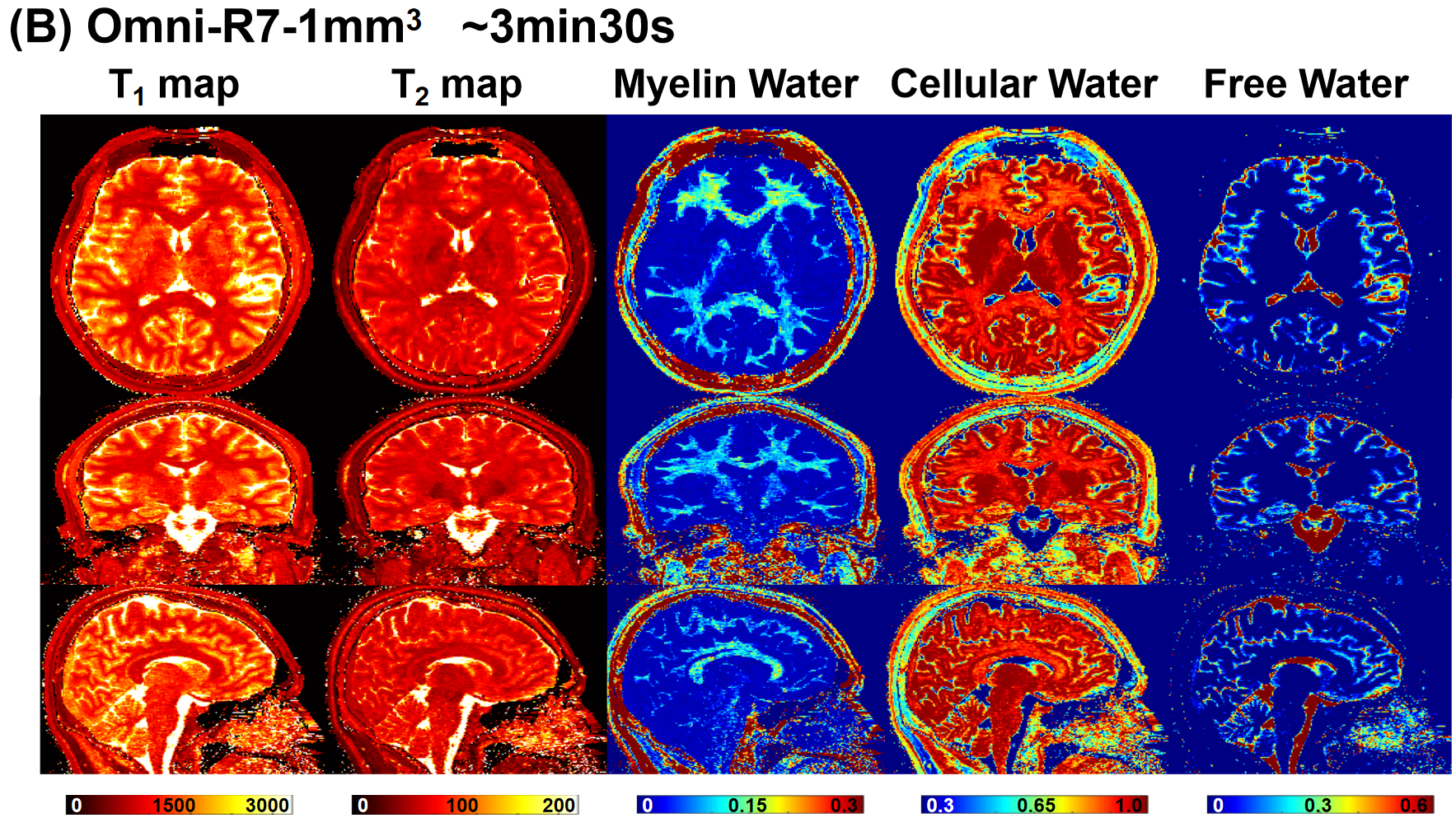}
\caption{Representative example of 3D brain tissue maps. Multi-planar reconstruction views of $T_1$, $T_2$, MWF, CWF, and FWF maps generated using the Omni-QALAS sequence. \textbf{(A)} Maps obtained with an acceleration factor 5. \textbf{(B)} Maps obtained with an acceleration factor 7.
}
\label{3d}
\end{figure*}

The $T_1$ and $T_2$ quantification results of the three sequences are shown in Figure~\ref{t1t2}. Subspace reconstruction takes approximately two and a half minutes per slice, resulting in a total reconstruction time of about 11 hours for the whole brain. To further compare the quantification results across different sequences, five regions of interest (ROIs) were selected, and the mean and standard deviation within these regions were calculated.

Omni-QALAS consistently shows a smaller standard deviation, particularly for $T_2$, indicating improved SNR. Specifically, the standard deviation values for Omni-QALAS across ROI-1 to ROI-5 were [2.4, 3.3, 7.5, 5.9, 93.6], corresponding to improvements of approximately 3.16-, 3.15-, 1.18-, 1.20-, and 1.18-fold compared to the QALAS sequence. Nevertheless, in ROI-5, the standard deviation remains relatively large even in Omni-QALAS. This can be attributed to the fact that the optimization process only considered the parameters of gray matter, white matter, and myelin water, without incorporating cerebrospinal fluid (CSF), which is characterized by long $T_1$ and $T_2$ values. As a result, the quantification accuracy for CSF was not effectively improved.


In addition, water fraction maps were also computed, as shown in Figure~\ref{mwf}. These were categorized into myelin water fraction, intra-/extra-cellular water fraction, and free water fraction. 

Figure~\ref{3d} shows representative multi-planar views of 3D T\textsubscript{1}, T\textsubscript{2}, myelin water, intra-/extra-cellular water, and free water tissue fraction maps of the brain acquired from a healthy volunteer using Omni-QALAS. Omni-R7-1mm\textsuperscript{3} achieved comparable mapping performance to Omni-R5-1mm\textsuperscript{3}, despite requiring only 3 minutes and 30 seconds of scan time, highlighting its improved acquisition efficiency.

Regarding water fraction mapping, the three sequences perform similarly in estimating the cellular and free water fractions. However, for the myelin water fraction, QALAS exhibits reduced contrast, potentially leading to underestimation of this compartment. In contrast, MWF-QALAS demonstrates improved delineation of myelin-related structures, and the Omni-QALAS sequence achieves the most detailed depiction. In regions where fine anatomical structures are visible in the reference map, QALAS fails to resolve them clearly, whereas the Omni-QALAS provides visibly enhanced structural detail and spatial definition. We computed the Shannon entropy\cite{Obuchowicz2020,Bayesian2017} within the region shown in Figure~\ref{mwf}, which was calculated as:

\[
H = -\sum_{i=1}^{L} p(i) \cdot \log_2 p(i)
\]

where \( p(i) \) represents the probability of MWF values falling into the \(i\)-th bin, and \(L\) is the total number of bins. Higher entropy values indicate increased heterogeneity and structural complexity within the selected ROI.

The computed entropy values were:
\[
H_{\text{QALAS}} = 2.49; \quad H_{\text{MWF}} = 2.68; \quad H_{\text{Omni}} = 2.78
\]

\begin{figure*}[h]
\centering
\includegraphics[width=1\linewidth]{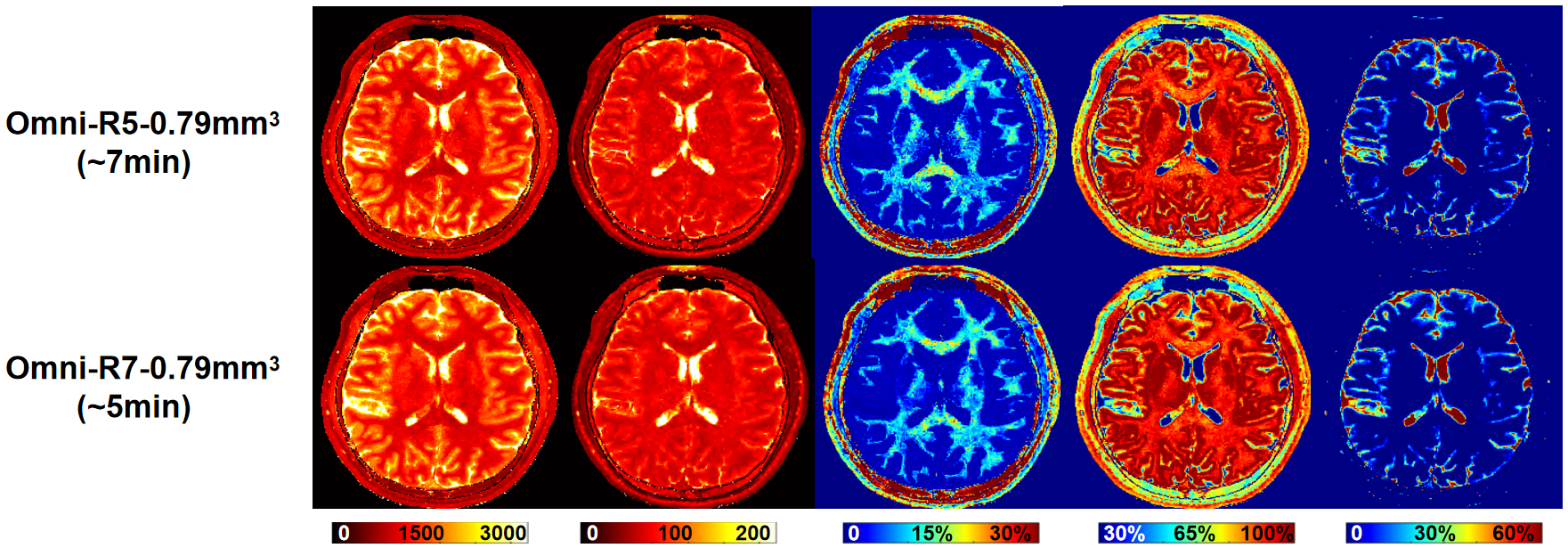}
\caption{The figure illustrates the performance of the Omni-QALAS sequence under different undersampling rates and spatial resolutions. The $T_1$ map, $T_2$ map, MWF, CWF, and FWF maps are presented. }
\label{R7}
\end{figure*}

Figure~\ref{R7} presents the quantitative results obtained from the higher resolution sequences. When using a higher spatial resolution of 0.79~mm\textsuperscript{3}, the resulting $T_1$ and $T_2$ maps exhibited increased noise levels, likely due to the reduced SNR associated with smaller voxel sizes.

\section{CONCLUSION and DISCUSSION}
In this study, we addressed the limitation of QALAS in capturing short \(T_2\) signals by improving the quantification performance of MWF-QALAS and reducing the number of readouts from six to five through CRB-based MLP optimization. The optimization incorporated tissue-specific objectives (gray matter, white matter, myelin water) and scan time constraints. The resulting sequence, Omni-QALAS, reduced the number of readouts from six to five, resulting in a scan time nearly one minute shorter than that of MWF-QALAS, while enhacing quantification accuracy.


Compared with the 3D ViSTa-MRF method proposed by Liao et al.~\cite{Liao2024}, which achieves 1.0~mm and 0.66~mm isotropic resolution in 5 and 15 minutes respectively, our proposed Omni-QALAS enables 1.0~mm resolution in a shorter scan time of 3 minutes and 30 seconds, and 0.79~mm resolution in 5 minutes. Compared to the original QALAS and MWF-QALAS, Omni-QALAS sequence improves short \(T_2\) estimation while maintaining or enhancing \(T_1\)/\(T_2\) accuracy.

Parameter settings during the optimization impact the outcome. For instance, the optimization range of the flip angle was set based on the work of Kvernby et al\cite{Kvernby2014}. We selected a range of 2° to 6° with a mean of 4°, as a wider flip angle range tends to compromise the accuracy of $T_1$ quantification. Additionally, we observed that the GAP tends to increase during optimization to promote $T_1$ recovery, thus a time penalty term was added to the loss function to discourage excessive scan time. The optimized TE values refer to the echo times of the T\textsubscript{2} preparation pulses, consisting of a long TE (70.5~ms) , and a short TE (20.9~ms).

However, Omni-QALAS showed suboptimal performance in the CSF region, as CSF (characterized by long $T_1$ and $T_2$) was not included among the targeted tissue types during optimization. If more accurate CSF quantification is desired, its $T_1$/$T_2$ values can be incorporated into the optimization framework.

For solving Equation~\ref{objective}, Lee et al.~\cite{Lee2019} and Arefeen et al.~\cite{Arefeen2023} used the Sequential Least Squares Quadratic Programming (SLSQP) implementation provided by SciPy\cite{SciPy} to optimize flip angles. However, we found it ineffective when applied to the full \( n \times \text{ETL} \) sequence. To address this, we adopt an MLP, which can learn complex nonlinear relationships and flexibly incorporate multi-objective losses, such as scan time constraints.

The proposed MLP-based optimization framework is equally applicable to the original QALAS sequence. As demonstrated in Figure~S5--S7, an optimized four‑readout QALAS outperforms the standard five‑readout implementation, achieving a 20\% scan‑time reduction and improved quantification accuracy. Moreover, this framework is also applicable to other imaging sequences, such as MR Fingerprinting\cite{Chen2019,Lancione2024}, offering a broad utility in MRI sequence optimization. This is because MR Fingerprinting similarly involves adjustable parameters such as FAs and repetition times (TRs) that can be optimized, and its signal model can be readily adapted to the MLP optimization framework.

Finally, subspace reconstruction~\cite{Jun2024} was essential for mitigating artifacts caused by varying FAs. It improves robustness, especially under undersampling R=7.

\bibliographystyle{abbrvnat}
\bibliography{main}
\end{document}